\documentclass{article}
\usepackage{amsfonts}
\usepackage{graphics}
\usepackage{epsfig}
\def\del{\partial}

\def\a{\alpha}

\def\t{\tau}

\def\Mgnn{M_{g,n}^{1^n}}

\def\Arc{\mathcal{ARC}}

\def\del{\partial}

\def\text{\mbox}

\def\A{\widetilde{Arc}}

\def\PB{\mathcal{P}({\mathcal B})}
\def\BV{\Delta}
\def\Ass{\mathcal A}

\newlength{\tskip}

\newcounter{thMM}
\newcounter{leMM}
\newcounter{deFF}
\newcounter{exMP}
\newenvironment{theorem}[1]{\refstepcounter{thMM}\trivlist
  \item[\hskip19pt{\sc #1~\arabic{thMM}.}]\it\hskip3pt}{\endtrivlist}

\title{Noncommutative aspects of open/closed strings via foliations}

\author{Ralph M. Kaufmann\\
Purdue University,
 Department of Mathematics\\
150 N. University Street, West Lafayette, IN 47907-2067\\
email: rkaufman@math.purdue.edu}

\date{Ferbruary 18, 2008}

\begin{document}

\maketitle

\begin{abstract}
We give a brief summary of algebraic aspects of string theory
arising in the noncommutative geometry setting of foliations called
string diagrammatics which we introduced jointly with Bob Penner. We
furthermore discuss how this gives rise to actions on the Hochschild
complex of a Frobenius algebra. We then explain how this leads to
new quantum chains for loop spaces and a stabilization in the
semi--simple case.
\end{abstract}

%\tableofcontents

\section{Introduction}
Over the years there have been various mathematical incarnations of
the operations of string theory. These come in many different
flavors as we recall in \S2. Here we will view moving strings as
naturally defining surfaces with partially measured foliations. On
this geometric/topological level the string interactions are
described by operadic structures. These are the mathematical
structures that describe  gluing operations where both the surfaces
and the foliations are glued. The result which is joint with Bob
Penner is a combinatorially defined open/closed CFT with partial
compactification \cite{KP}.

This approach is very powerful as it gives the right phenomenology.
First on the degree zero level, that is restricting to connected
components, one obtains a new proof of the axioms of open/closed
TFT. Secondly we obtain the correct BV formalism in the closed
sector and interesting new behavior in the open/closed sector.
Passing to the homology or operator level, we were able to give an
action on the Hochschild co--chains of a Frobenius algebra which had
been expected from String Topology, $D$--branes and purely on
mathematical grounds \cite{hoch}. Finally, we give applications in
the semi--simple case by defining a quantum loop space and
considering stabilizations of the moduli spaces.

\section*{Acknowledgements}
It is a pleasure to thank our collaborator Bob Penner.
We also thank the organizers of the ISQS XVI for providing
this wonderful forum.

\section{Moving Strings as a Noncommutative Geometry of  Measured Foliations}
\label{movinpar}

The basic picture we have in mind is that strings are either
parameterized circles or intervals. As such they have a measure
sitting on them and special points, namely $0$ in the closed case or
the endpoints of the interval in the open case. Now as the strings
move, split and recombine they sweep out a surface. Depending on our
point of view, we could stop here and we would be considering TFT.
We could also include more data as discussed below. The second
fundamental aspect of this {\em a posteriori} actually
not--so--na\"ive picture is that we can let the string system run
for a bit, then stop and then let it run again. The final result
will be a surface with has been glued from the surface swept out
during the first time period and the surface swept out during the
second. This is what gives rise to the operadic structure.

In order to be more precise, we need to fix the geometry we are talking about.
In the table below, we fix the data we are gluing ---which will always
be surfaces together with extra data--- and specify the theory we are encoding
mathematically. In order to simplify things we will for the moment restrict to the closed sector. This means there will be only one distinguished point per
boundary component.

\begin{tabular}{l|l|l}
Geometry&data (roughly)&Theory\\
\hline&&\\[\tskip]
Topological surfaces $\Sigma$ w/ boundary $\del\Sigma$/ &$(\Sigma, \del \Sigma)$&TFT\\
Cobordisms thereof&&\\
\hline&&\\[\tskip]
Surface $\Sigma$ w/  boundary $\del\Sigma$
  &$(\Sigma, \del \Sigma, [g])$&CFT\\
and conformal structure $[g]$/&&\\
``Segal operad/category''/&&\\
Open moduli space $M_{g,n}$ operad&&\\
\hline&&\\[\tskip]
Complex curve $C$ w/   &$(C, p_1, \dots, p_n)$&CohFT\\
marked points $p_i$&&GW invariants\\
 nodes possible $C\in\bar{M}_{g,n}$&&\\
\hline&&\\[\tskip]
Surface $\Sigma$ w/ boundary $\del\Sigma$&
$(\Sigma, \del \Sigma, p_i\in \del_i\Sigma, [\a])$&Combinatorial\\
marked points $p_i\in \del_i\Sigma$&& CFT\\
and class of Foliations $[\alpha]$&&\\
\end{tabular}

We will be concerned with the last entry.

%\subsubsection{open/closed}
In order do the open/closed version we will have to add more points
on the boundary. These points correspond to the ends of open strings
and are hence labelled by a set of $D$--brane labels chosen from a
set of elementary labels ${\mathcal B}$.  To give the gluing
structure we will pass to the power set ${\mathcal P}({\mathcal B})$
as the labelling set for the points. The label $\emptyset$ will mean
``closed string''. If a point has a collection of labels, we think
of this as the intersections of relevant branes. It is allowed that
these can all be empty in a realization.

\section{String diagrammatics on the space, chain and operator level}

%\subsection{Strings as Foliations}
Adopting the point of view that the strings have measures, we see
that the surface they sweep out actually has a foliation, which has
a transversal measure, see Figure 1.
%\ref{figone}
 We can think of each point of the string as giving a leaf of
this foliation. The strings themselves will also give a foliation
transversal to the previous one. Each string has a measure and hence
the first foliation has a transversal measure. Going on, we can
``squeeze'' the leaves together like a curtain to fit into bands on
the surface. These bands will now end on an interval between two
marked points -- called windows.
 As the bands also come with a transversal measure,
  they have a width.
Thus the whole data can be thought of as combinatorially given by a
graph on the surface --one edge for each band-- whose edges are
labelled by positive real numbers that indicate the width. We will
enlarge the picture by allowing the graphs to degenerate. This means
that the underlying ribbon graph can have a different topological
type from the surface it lies on, see \cite{KP,KLP} for details and
examples. We call a foliation admissible if its graph as a graph on
the surface has no parallel (homotopic) edges and none of the edges
are parallel to a window, where in the homotopy the endpoints are
not allowed to pass each other or the marked points. We also exclude
the cases in which an edge is parallel to a part of the boundary
including a point marked with $\emptyset$. Finally, we require that
there is at least one edge.

\subsection{Basic descriptions of the interactions}

Simply looking at the string interaction, we  stay on the level
spaces of foliations of surfaces. Here the string interaction is
captured by gluing the surfaces along the interacting
strings/boundary components as depicted in Figure 1.
%\ref{figone}.
The point of view of \cite{KP} is that we only glue if the widths
are the same. This corresponds to the dynamic picture mentioned
above. The important thing to notice is that in the open gluing, the
marked points become punctures, while in the closed gluing they are
erased after the gluing. It is this fundamental difference which
leads to the correct algebra of interactions.

We can go one step further and consider parameterized
families of these operations.
What this basically means is that we consider varying weights on the bands.
Here the topology of the space is given
by varying the weights and erasing a band as its width tends to zero.
Just as the gluing space level
operations above there are family gluing operations as well.

With a lot of work, \cite{KP} we could show that there are operations
induced on the homology level as well. What this means is that
we obtain vector spaces of operators. These are graded by their
degree which is the number of parameters. These operators then
define a version of open/closed CFT, see below. The lowest degree $0$
corresponds to the connected components and hence to TFT. Using this
description gives a new proof for minimal axioms of open/closed TFT
\cite{CL}.

\subsection{A mathematical description of the structure}

We will call $\A(n,m)$ the space of classes of admissible foliations
under orientation-preserving homeomorphisms  of brane labelled
surfaces with $n$ active closed window and $m$ active open windows.
Here we call a window ``open'' if both endpoints (which may
coincide) are labelled by a nonempty $D$--brane label and a window
``closed'' if it is on a boundary component with one marked point
that is marked by $\emptyset$. Being active means that the window is
hit by leaves of the foliation.

\begin{theorem}{Theorem}
The spaces $\A(n,m)$ of surfaces with partially
measured foliations form an ${\mathbb R}$ graded
 C/O--structure.
\end{theorem}

 This is a new type of operad--like structure defined in \cite{KP}.
Being a C/O structure
roughly means that we can associatively glue a closed window to a closed one or
an open window to an open one
under the condition that the weights agree. Here self--gluings of
surfaces are allowed, that is the two windows that are glued
can either lie on two different surfaces or on the same surface.
 One of the really surprising results
which requires hard work is:

\begin{theorem}{Theorem}
The operations on the space level induce the structure of a
bi--modular bi--operad on the homology level.
\end{theorem}

This is the gold standard for mathematical descriptions of CFT and
we were not able to show this using the previous approach of
\cite{KLP} where we allowed more gluings on the space level using an
overall projective scaling of the weights. This in turn however
prevented us from defining self--gluing. Of course on the homology
level the operations of \cite{KLP} and \cite{KP} agree. Another
important result is that we can even lift the  operations above to
the chain level, which is important for applications to String
Topology. This fact is  actually needed in the proof of the Theorem
above which consists of a careful analysis of flows of foliations on
the respective surfaces. It should be remarked that on the homology
level
 there is no condition on
weights anymore.

\subsection{Operators and Relations}

The homology is linear and hence we can look for (operadic)
algebra representations of it. I.e.\ each  element of homology defines
an operator in such a way that the relations are
preserved under gluing. This means roughly that gluing two
families representing
homology classes corresponds to concatenating the relevant
operators. Notice that homology is graded and these degrees should
be preserved. A degree $k$ class can be represented by a
$k$--dimensional family and this should correspond to an operator of
degree $k$. The degree $0$ classes correspond to points up to
homotopy and  are therefore represented by the connected components.

Disregarding the data of the parameterized families of graphs, each
surface  with marked points on the boundary and a foliation can be
decomposed into bi-gons, triangles, annuli with one point on each of
the boundaries, once punctured annuli of the same type, discs with
two punctures and one marked point on the boundary and pairs of
pants with one marked point on each boundary. The surfaces without
punctures are exactly the surfaces of Figure 2. Moreover the
indecomposable families of degree $0$ and $1$ without punctures are
exactly the ones depicted in Figure 2.

Different decompositions of surfaces with families of foliations
give rise to relations. Different pieces can glue together to form
the same surface and therefore the composition of the respective
operators should agree.
 In this way one obtains all the expected relations in
degree $0$. The relations of degree zero involving the open sector
and the open/closed interaction are for example depicted in Figure
3.
For the standard Frobenius relations on the closed sector and the
relations of degree $1$ which establish that $\Delta$ is a BV
operator that is compatible with a natural Gerstenhaber bracket we
refer the reader to \cite{KP}.

\begin{theorem}{Theorem}
An algebra over the  modular bi-operad $H_*(\coprod_{n,m}\A(n,m))$
is a pair of vector spaces $(C,\Ass)$ with the following properties:
$C$ is a commutative Frobenius BV algebra $(C,m,m^*,\BV)$, and
$\Ass=\bigoplus_{(A,B)\in {\PB\times \PB}}\Ass_{AB}$  is a
$\PB$-colored Frobenius algebra (see e.g., \cite{LP} for the full
list of axioms of such a structure). In particular, there are
multiplications $m^{ABC}:\Ass_{AB} \otimes \Ass_{BC}\to \Ass_{AC}$
and a non-degenerate metric on $\Ass$ which makes each $\Ass_{AA}$
into a Frobenius algebra.

Furthermore, there are morphisms $i^{A}:C\rightarrow \Ass_{AA}$
which satisfy the following equations: letting $i^*$ denote the dual
of $i$, $\t_{12}$ the morphism permuting two tensor factors, and
letting $A,B$ be arbitrary non-empty brane-labels, we have

\begin{eqnarray}
&i^B\circ {i^A}^* =  m_B \circ \t_{12}\circ m^*_A& (\text{Cardy})\\
&i^A(C)~{\rm is~central~in}~{\mathcal A}_A &(\text{Center})\\
& i^A\circ \BV \circ {i^B}^*=0& (\text{BV vanishing})
\end{eqnarray}
These constitute a spanning set of operators and a complete set of
independent relations in degree zero.  All operations of all degrees
supported on indecomposable surfaces are generated by the degree
zero operators and $\Delta$.
\end{theorem}
Here the intriguing equation is the BV vanishing. This is a new
feature, which has to do with the partial compactification.

Using the results of transitivity of the four moves on
decompositions of constant foliations, see Figure 3,
and the theorem above we also obtain the restriction to an
open/closed TFT:
\begin{theorem}{Theorem}
An algebra over the degree $0$ part of the operad, that is an
open/closed TFT is precisely given by the data  $(C,\Ass)$ which
have the following properties: $C$ is a commutative Frobenius
algebra, and $\Ass=\bigoplus_{(A,B)\in {\PB\times \PB}}\Ass_{AB}$ is
a $\PB$-colored Frobenius algebra which satisfies the Cardy and
Center equation.
\end{theorem}

Hence, as the degree $0$ part, we recover the Cardy/Lewellen
axiomatic picture (\cite{CL}, see also \cite{LP}) from the point of
view of strings yielding foliations. This includes the
non--commutative Frobenius algebras of the open string sector and
the easy description of the Cardy equation of Figure 2.

\subsection{Moduli space/CFT} In this section, we will restrict
ourselves to the closed string sector.  As we stated above, we
augmented the admissible foliations, by allowing the graphs to be
degenerate ---  that is of a ``smaller'' topological type than the
surface. To be precise a graph of a foliation is called
quasi--filling if the complementary regions are either polygons or
once punctured polygons. Let $\Arc_{\#}(g,n,s)$ denote the space of
quasi--filling graphs of closed sector surfaces of genus $g$ with
$s$ punctures and $n$ marked points, that is surfaces whose
boundaries each only have one marked point labelled by $\emptyset$.

\begin{theorem}{Theorem}
\cite{hoch,P2} $\Arc_{\#}(g,n,s)$ is homotopy equivalent to the
decorated moduli space of surfaces of genus $g$ with $s$ punctures
and $n$ marked points. In particular, if there are no punctures
$\Arc_{\#}(g,n,0)\cong M_{g,n}^{1^n}$ that is it is isomorphic to
the moduli space of genus $g$ curves with $n$ punctures and one
tangent vector at each puncture.
\end{theorem}

In other words, we recover moduli space and hence restricting the
operad structure to this subspace we obtain a combinatorial version
of CFT. This is however a bit subtle, since on the topological level
the statement is true only {\em cum grano salis}. The mathematically
correct statement is

\begin{theorem}{Theorem}
Using the gluing of \cite{KLP} the subspaces $\Arc_{\#}(g,n,0)$ form
a rational (i.e.\ densely defined) cyclic operad which induces a
cyclic operad structure on the relative chain complex of open cells indexed by
marked ribbon graphs (see \cite{hoch} for full details).
\end{theorem}

\section{Actions on Hochschild}

\subsection{Motivation} There are three sources of motivation to look
for actions of a chain model of moduli space. Perhaps the most
intriguing come from using the logic of Kontsevich-Kapustin-Rozansky
\cite{KR}, which we can rephrase as follows. If the closed string
states are thought of as deformations of the open string states and
the open string states are represented by a category of $D$-branes,
then the closed strings should be elements of the Hochschild
co--chains of the endomorphism algebra of this category. Now
thinking on the worldsheet, we can insert closed string states. That
is, for a world sheet, we should get a correlator by inserting, say
$n$ closed string states. This is what we will have done, if one
simplifies to a space filling $D$-brane and twists to a TCFT.

The second motivation is from String Topology \cite{CS}, where
surfaces should act on the homology of the loop space of a manifold.
Now it is well established that if the manifold is simply connected,
then the homology of the loop space is calculated by the Hochschild
complex of co-chains of the manifold. Going one step in the spectral
sequence and supposing that the manifold is compact and hence has
Poincar\'e duality, we again expect a cell level action of moduli
space.

Lastly just looking at Deligne's conjecture and its generalizations
(see \cite{hoch} for a
full list of references and proofs thereof) we are motivated to
include all surfaces.

\subsection{Results on actions}
As expected there is indeed such an action on the Hochschild co--chains,
which can be understood as an action of the discretization of the foliations.

\begin{theorem}{Theorem}
There is  an action on the Hochschild co--chains of a Frobenius
algebra by the relevant chain complex of ribbon graphs which
calculates the co--homology of moduli  space. Restricting to a
particular subpart and partially compactifying, we obtain an action
which is the one of String Topology of Chas and Sullivan (possibly
up to lower order terms) \cite{hoch}.
\end{theorem}

\subsection{Some details on the action relevant to the further discussion}
Fix a commutative unital Frobenius algebra $A$ with multiplication
$\mu$ and paring $\langle \; ,\; \rangle$. Set $\int a:=\langle
a,1\rangle$ and let $e$ be the Euler element of $A$ that is
$e=\mu\Delta(1)$ where $\Delta$  is the adjoint of $\mu$. Using
dualization, we need to define correlators: $ \langle
\phi_1,\dots,\phi_n\rangle_{\Sigma_{g,n},\Gamma}$ for any cell given
by a surface $\Sigma_{g,n}$ with boundary and marked points as
before and a ribbon graph $\Gamma$ representing a foliation $[\a]$
with varying weights.
 Here we think of $\phi_i\in TA \simeq C^*(A,A)$
where $TA$ is the tensor algebra. The action is now roughly given as
follows (full details are given in \cite{hoch}): (1) Duplicate edges
so that the number of incoming edges at the vertex $i=deg(\phi_i)$.
We sum over all possibilities to do this, if this is not possible
then the operation is zero. (2) Assume the $\phi_i$ are pure
tensors. Pull apart the edges and decorate the pieces of the
boundary with the elements of $\phi$. Cut along all the edges of the
graph and call the set of disjoint pieces of surface $P$. Let $I(p)$
be the index set of the components $a_j$ of the $\phi_i$ decorating
edges belonging to a piece $p\in P$ and let $\chi(p)$ be the Euler
characteristic of the surface $p$. Notice that the pieces $p$ which
possibly have non--trivial topology are a subset of the pieces of
the surfaces, that are obtained by cutting along the original edges
before duplication. (3)~Set $\langle
\phi_1,\dots,\phi_n\rangle_{\Sigma_{g,n},\Gamma}:=\prod_{p\in P}
\langle\phi_1,\dots,\phi_n\rangle_p$ where
\begin{equation}
\label{coreq} \langle\phi_1,\dots,\phi_n\rangle_p=\int \prod_{i\in
I(p)}a_i e^{-\chi(p)+1}
\end{equation}

\section{Stabilization, the semi--simple case and a quantum loop space}

We wish to point out that the topology of the surface pieces $p$
enters only through the factor $e^{-\chi(p)+1}$. This factor is
invertible for $\chi(p)\neq 1$ precisely if the algebra $A$ is
semi--simple. Moreover  if $A$ is semi-simple with idempotents
$e_i$, set $\lambda_i=\int e_i$ then  $e=1$ if  all $\lambda_i=1$.
We will call such an algebra a normalized semi--simple Frobenius
algebra. All semi--simple Frobenius algebras can be obtained from a
normalized one by scaling the metric. Moreover, any semi--simple
finite dimensional algebra can be endowed with a metric that makes
it into a normalized semi--simple Frobenius algebra.

For such an algebra the action of a cell $(\Sigma,\Gamma)$ equals to
the action of the surface where the pieces $p$ are replaced by
discs. We call this operation stabilization.  Here the boundary
pieces of $p$ are cut and glued to an $S^1$ in a fixed fashion,
which keeps the relative order of the pieces intact.

\begin{theorem}{Theorem}
For a normalized semi--simple Frobenius algebra the action factors
through the stabilization. Moreover the stabilized surfaces form an
operad and contain an $E_{\infty}$ suboperad. Hence the Hochschild
co-chains of a normalized semi--simple Frobenius algebra have the
structure of an $E_{\infty}$ algebra.
\end{theorem}

The first part of this is immediate. For the  second we
 construct the stabilization via a colimit whose maps actually
 add topology to the pieces $p$ and explicitly
 give the $E_{\infty}$
 operations as the geometrization of the operations of
\cite{MS}. If $A$ is not semi--simple the action does not pass
through the stabilization. We can however  flow the metric to a
normalized one which has the effect of changing the
co--multiplication. Also, in the semi--simple case the operations
are shifted from the stabilized ones by invertible elements.

\subsection{Quantum chains on the Loop space and the quantum string bracket}
It is well known that  there is a cyclic model for the chains of
loop space of a simply connected manifold $M$ given by
$C^*(S^*(M),S_*(M))$. Calculating the homology we are lead to the
$E_1$ term of the spectral sequence which is isomorphic to
$C^*(A,A)$, with $A=H^*(M)$. Now say $M$ is a smooth projective
variety, we can quantum deform $A$ to $A_q:=H_q^*(M)$, where $H_q$
is the quantum cohomology at the point $q$. This gives us a model
for {\em quantum chains on the loop space}: $C^*(A_q,A_q)$. As such
we have the operations above and in particular the string bracket
induced by the Chas--Sullivan PROP.

Now if $A_q$ is semi--simple  ---this is expected generically if
 $M$ is a Fano variety that has a system of
exceptional sheaves of appropriate length, e.g.\ ${\mathbb P}^n$---
then we can actually use gravitational descendants to flow to a
normalized semi--simple Frobenius algebra \cite{KMZ,T}

\begin{theorem}{Theorem}
Given a semi--simple point, we can flow to a normalized point
(essentially by coupling to gravity) at which the quantum chains on
the loop space are an $E_{\infty}$ algebra. In this situation the
quantum deformed string topology bracket
 vanishes.
\end{theorem}

\subsection{Connections to  family field theories and Mumford--Morita--Miller
Classes} The correlators on $V=TA \simeq C^*(A,A)$ for a Frobenius
algebra define chain level family field theories with partition
functions $Z_{g,n}\in Hom(V^{\otimes n},Hom(C^*(\Mgnn),k))$. One
result is that when $A$ is semi--simple, the operations on the
degenerate higher genus Penner--boundary actually come from lower
genus moduli space via an invertible morphism. We expect to show
that we can identify our stabilization with that of Tilmann, Segal
and Madsen \cite{MT} and  prove that the partition function is
actually fixed by its genus zero contribution. For this we can pull
back co-chains along our stabilization maps of the colimit which
induces push--forward on the $Hom$--duals. This means that we can
effectively push the operations to infinite genus. If we have an
identification with the usual stabilization we can apply the Mumford
conjecture proved in \cite{MW}.  The upshot will be a classification
result for these theories in terms of $\kappa$ classes along the
lines of \cite{KMZ,T}.

In order to compare to GW--invariants, we need to  ``go to the
boundary'' of moduli space in the Deligne--Mumford sense.  One
approach   is using frames of graphs  \'a la Fulton--MacPherson (see also \cite{Penner})
  another is to
cut the surface along a curve that does not intersect any arcs and
then contracting these curves to points. This essentially yields
Kontsevich's compactification. One indication that this approach is
natural is given by the observations in Appendix B of \cite{KP}. The
next step is then to lift the $Z_{g,n}$ from the Kontsevich
compactification to the DM compactification. Going beyond this we
wish to point out that the underlying algebra $TA_q$ is the algebra
governing the quantum cohomology of the symmetric products
\cite{sq}. These in turn define the genus $g$ GW invariants as
discussed in \cite{costello}. This would
provide another approach to Teleman's classification result \cite{T}
and give further insight to Givental's ``loop space approach''
\cite{Giv}.

\section{Conclusion and Outlook}
We have discussed how to regard moving strings as providing
surfaces with foliations. Using this approach, we
 have recovered the axiomatics of
open/closed TFT, given a new combinatorial CFT and a partial
compactification of it which relates to/defines string topology
operations and lends itself to give
actions on the Hochschild complex of a Frobenius algebra.
We also obtained a partial list of axioms for the CFT
and partially compactified CFT. In the future we hope that we can
get a full axiomatic description.

Furthermore, we were able to give a chain level action of the
relevant structures generalizing Deligne's conjectures. Along the
same lines of reasoning it should be possible to get open/closed
operations on the Hochschild complex of a category or a pair of
algebras in the space filling $D$--brane situation, say in a
Landau--Ginzburg model.

Interesting future directions include the study of mapping spaces to
varieties/orbifolds. An even more ambitious project is to try to
connect our constructions to the Chiral deRham complex or other
bundle theories.

\begin{figure}[h]
\label{figone}
\includegraphics[height=202pt]{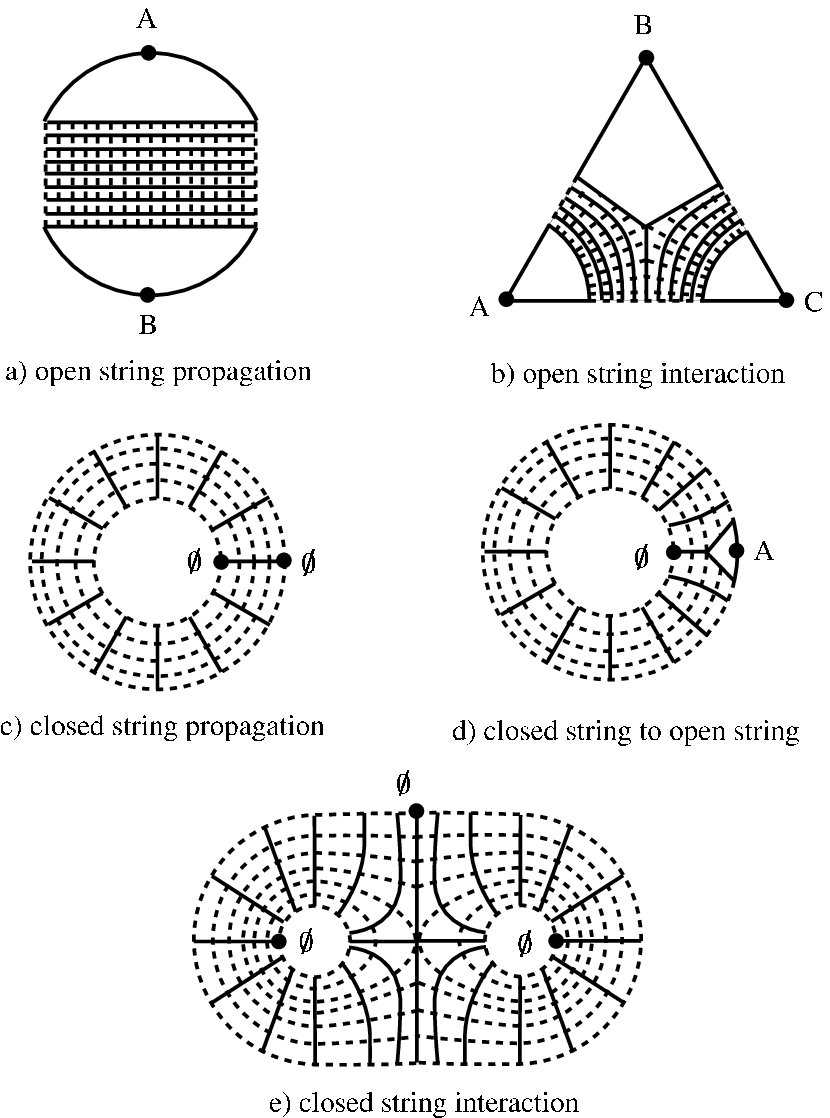}\hfill
\includegraphics[height=202pt]{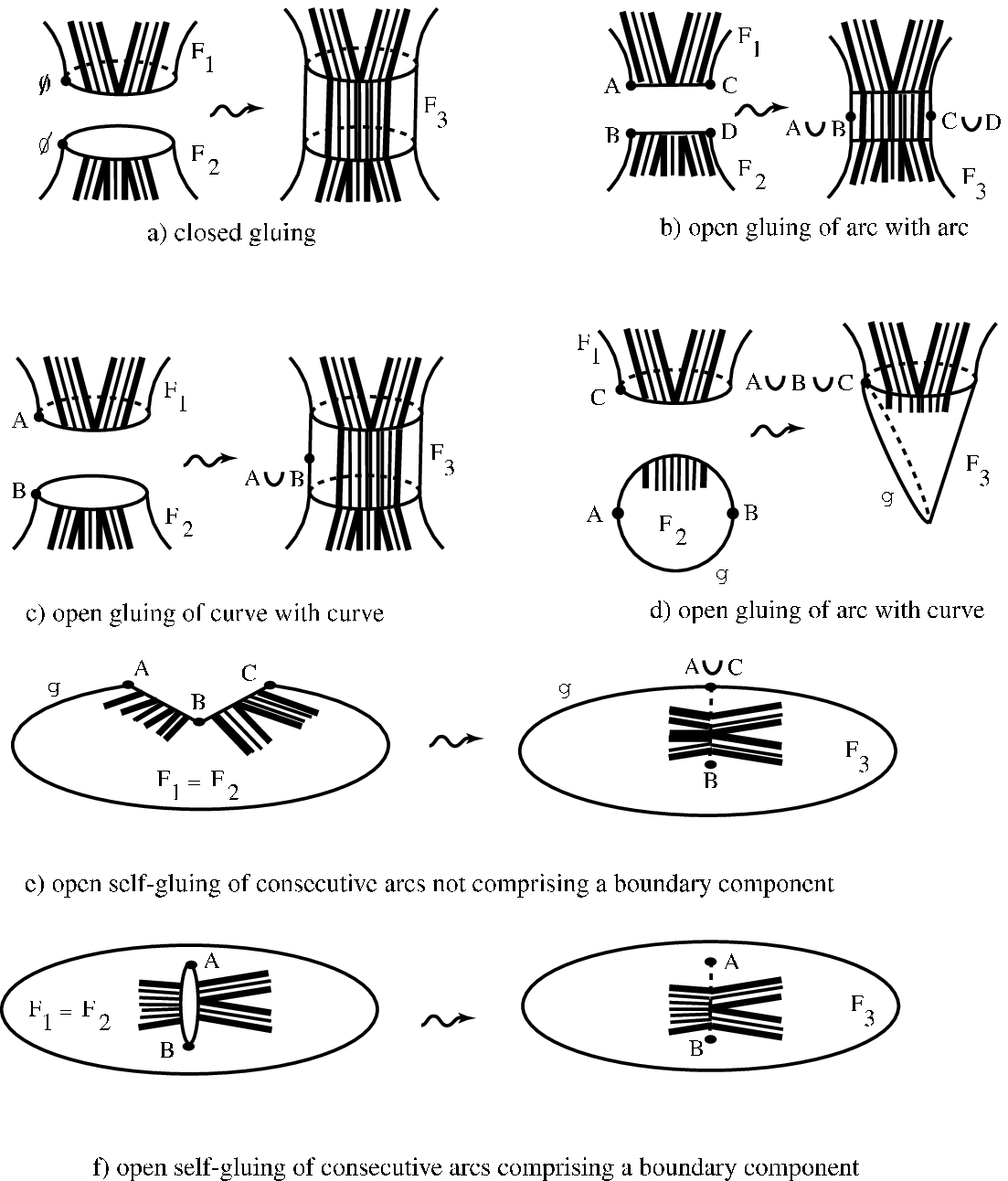}
\caption{Moving Strings as foliations and Glueings}
\end{figure}

\vfill

\begin{figure}[h]
\label{figtwo}
\includegraphics[viewport=0 245 253 385, height=120pt, clip]{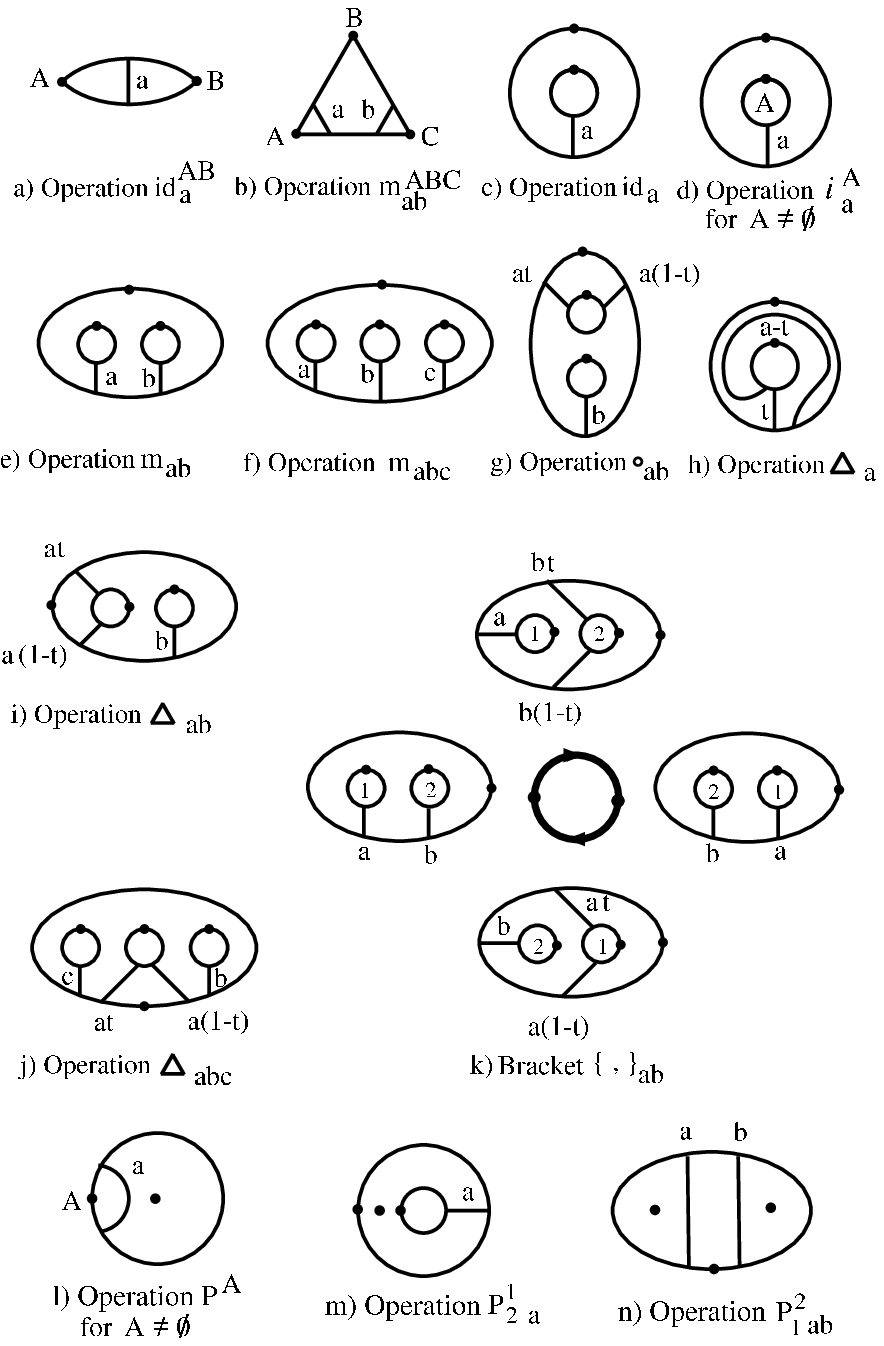}
\hfill\includegraphics[viewport=0 71 245 233, height=120pt,
clip]{stropfigops.eps} \caption{Operations}
\end{figure}

\vfill

\begin{figure}[h]
\label{figthree}
\includegraphics[height=115pt]{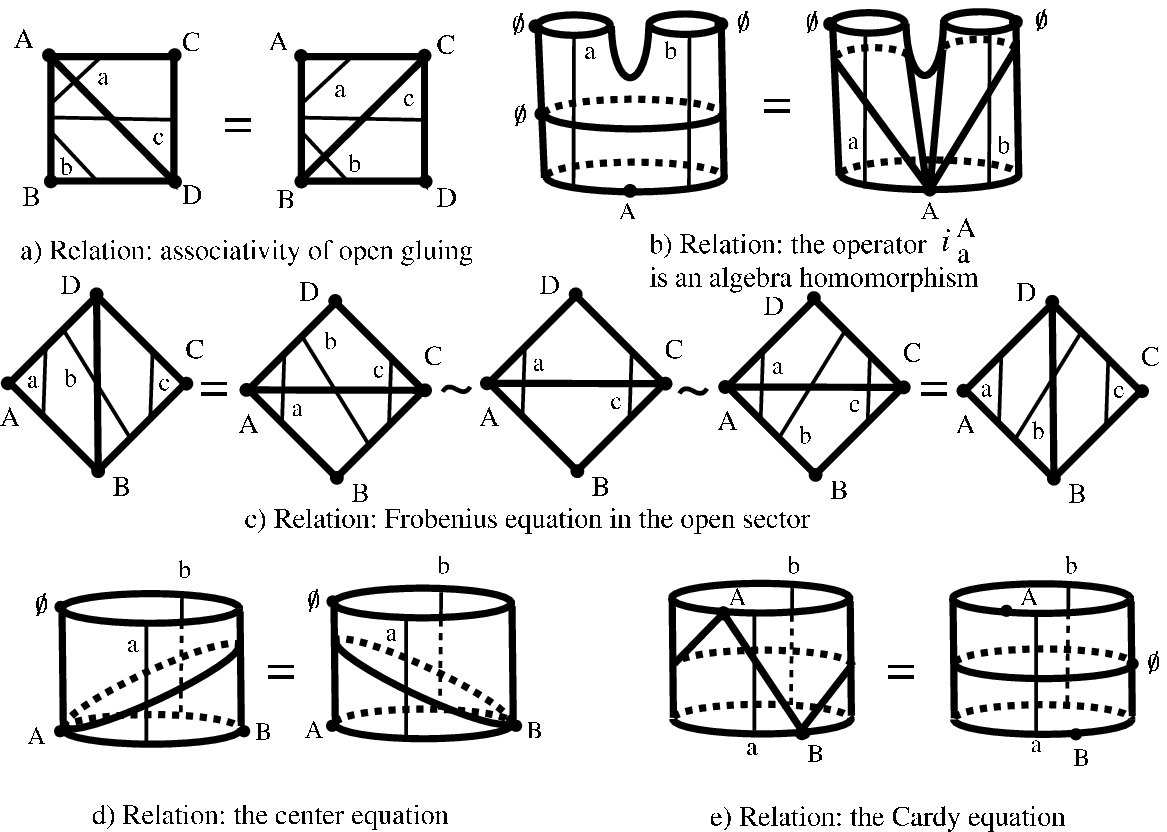} \hfill
\includegraphics[height=115pt]{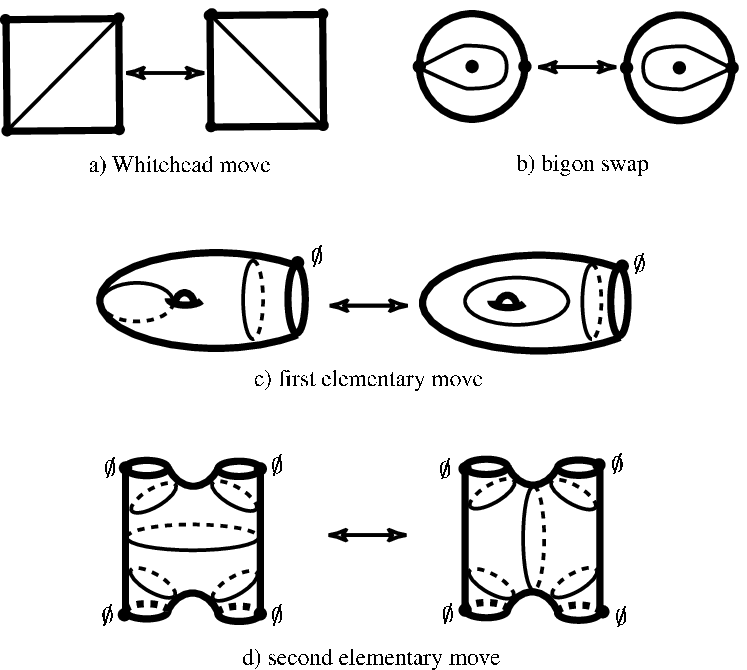} \hfill
 \caption{Moves and Relations}
\end{figure}

\end{document}